\documentclass[10pt,conference,a4paper]{IEEEtran}

\IEEEoverridecommandlockouts

\pdfpageattr{/Group << /S /Transparency /I true /CS /DeviceRGB >>}

\usepackage[utf8]{inputenc}
\usepackage[T1]{fontenc}
\usepackage{textcomp}
\usepackage{microtype}
\usepackage{calc}
\usepackage[normalem]{ulem}

\usepackage{lipsum}

\usepackage[range-phrase=--,per-mode=symbol-or-fraction,binary-units=true,range-units=single,list-units=single,detect-all,forbid-literal-units=true]{siunitx}
\AtBeginDocument{
\DeclareSIUnit\bit{bit}
\DeclareSIUnit\byte{Byte}
\DeclareSIUnit\decibelm{dBm}
\DeclareSIUnit\vehicle{veh}
}

\usepackage{silence}
\usepackage{csquotes}

\usepackage{amsmath}
\usepackage{amssymb}
\usepackage{amsfonts}
\usepackage{amsthm}

\usepackage{mathtools}

\usepackage{booktabs}
\usepackage{url}

\usepackage[inline]{enumitem}

\usepackage[caption=false,font=footnotesize]{subfig}
\usepackage[pdftex]{graphicx}
\DeclareGraphicsExtensions{.pdf,.png,.jpg}
\usepackage{tikz}
\usetikzlibrary{arrows}
\usetikzlibrary{calc}
\usetikzlibrary{chains}
\usetikzlibrary{scopes}
\usepackage{makecell}

\usepackage[american]{babel}
\usepackage{hyphenat}

\usepackage{algorithm}
\usepackage{algpseudocode}

\usepackage[capitalize,noabbrev,nameinlink]{cleveref}

\RequirePackage{xstring}
\RequirePackage{xparse}
\RequirePackage[]{acro}
\makeatletter
\@ifpackagelater{acro}{2019/02/17}{
	\NewDocumentCommand\acrodef{mO{#1}mG{}}{\DeclareAcronym{#1}{short={#2}, long={#3}, foreign-plural={}, #4}}
}{
	\NewDocumentCommand\acrodef{mO{#1}mG{}}{\DeclareAcronym{#1}{short={#2}, long={#3}, #4}}
}
\makeatother

\acrodef{CPU}{Central Processing Unit}
\acrodef{FCFS}{First Come First Served}
\acrodef{GNCU}{Google Normalized Computing Unit}
\acrodef{GNMU}{Google Normalized Memory Unit}
\acrodef{HPC}{high-performance computing}
\acrodef{JBT}{Join Below Threshold}
\acrodef{JIQ}{Join Idle Queue}
\acrodef{JSQ}{Join Shortest Queue}
\acrodef{MJS}{Mean Job Slowdown}
\acrodef{LWL}{Least Work Left}
\acrodef{MRT}{Mean Response Time}
\acrodef{RAM}{Random Access Memory}
\acrodef{RR}{Round Robin}
\acrodef{SITA}{Size Interval Task Assignment}
\acrodef{SRPT}{Shortest Remaining Processing Time}
\acrodef{SD}{Slowdown}

\usepackage[color]{changebar}
\def\todoCtd#1{%
	TODO: #1%
	\ifx&#1&.\fi%
	\endgroup
	\cbend
	\relax
}

\NewDocumentCommand\IEEE{ s m >{\SplitArgument{4}{/}}d[] }{%
    \IfBooleanTF{#1}{}{IEEE\,}
    \nolinebreak[2]
    #2%
    \IfNoValueTF{#3}{%
    }{%
        \sommerIEEELettersSlashed#3%
    }%
}
\newcommand{\sommerIEEELettersSlashed}[5]{%
    \IfNoValueTF{#2}{%
    }{%
        \nolinebreak[3]
    }%
	#1%
	\IfNoValueTF{#2}{}{/#2}%
	\IfNoValueTF{#3}{}{/#3}%
	\IfNoValueTF{#4}{}{/#4}%
	\IfNoValueTF{#5}{}{/#5}%
}

\begin{document}

\title{The Merit of Simple Policies: Buying Performance With Parallelism and System Architecture\thanks{This work was partially supported by the European Union under the Italian National Recovery and Resilience Plan (NRRP) of NextGenerationEU, in partnership with Telecommunications of the Future (PE0000001 - program "RESTART").}\\
\thanks{This is a preprint of the paper accepted to the 7th International Workshop on Intelligent Cloud Computing and Networking (ICCN) at IEEE INFOCOM 2025.}
}

\author{\IEEEauthorblockN{Mert Yildiz, Alexey Rolich, Andrea Baiocchi}
\IEEEauthorblockA{Dept. of Information Engineering, Electronics and Telecommunications (DIET), University of Rome Sapienza, Italy \\
\small{
        \texttt{
         mert.yildiz@uniroma1.it
        }%
        \texttt{
         alexey.rolich@uniroma1.it
        }
        \texttt{
         andrea.baiocchi@uniroma1.it
        }
    }
}

}

\maketitle

\begin{abstract}
\nohyphens{%
While scheduling and dispatching of computational workloads is a well-investigated subject, only recently has Google provided publicly a vast high-resolution measurement dataset of its cloud workloads.
We revisit dispatching and scheduling algorithms fed by traffic workloads derived from those measurements.
The main finding is that mean job response time attains a minimum as the number of servers of the computing cluster is varied, under the constraint that the overall computational budget is kept constant.
Moreover, simple policies, such as Join Idle Queue, appear to attain the same performance as more complex, size-based policies for suitably high degrees of parallelism.
Further, better performance, definitely outperforming size-based dispatching policies, is obtained by using multi-stage server clusters, even using very simple policies such as Round Robin.
The takeaway is that parallelism and architecture of computing systems might be powerful knobs to control performance, even more than policies, under realistic workload traffic.
}
\end{abstract}


\acresetall
\IEEEpeerreviewmaketitle

%

\section{Introduction}
\label{sec:intro}

The increasing demand for data-driven applications and the complexity of distributed computing environments have made job scheduling and dispatching in large-scale cluster systems a critical area of study. 
These clusters, which support cloud computing and high-performance computing infrastructures, need to balance system utilization, minimize response times, and ensure fair resource distribution.
The challenge is further complicated by the variability of workloads, which range from short, bursty tasks to long, resource-intensive jobs.

Enhancing the performance of large-scale clusters is closely tied to the efficiency of the dispatching and scheduling algorithms employed. 
Dispatching consists of assigning tasks to servers, while scheduling determines the sequence in which tasks are executed within each server. 
Both processes influence key performance metrics such as mean response time, system throughput, and load balancing. 
Traditional dispatching algorithms like \ac{RR} and scheduling algorithms like \ac{FCFS}  are simple, but often ineffective in environments with heterogeneous tasks, where job sizes and resource requirements can vary significantly.

To overcome these limitations, advanced dispatching strategies such as \ac{JSQ}, \ac{JBT}, \ac{SITA}  and \ac{LWL} , along with scheduling algorithms like \ac{SRPT}, have been introduced to enhance load balancing and reduce job completion times. 
It is known that in the case of non size-aware policies \ac{JSQ} is heavy-traffic delay-optimal \cite[Ch. 10]{SrikantYing2014}, as well as \ac{JBT} \cite{Zhou2017}.
As for size-based dispatching policies, it is known that there is no definite winner between \ac{SITA} and \ac{LWL} \cite{HarcholBalter2009}, yet they are both superior to other dispatching policies under the assumption of \ac{FCFS} servers. However, under heavy traffic, authors in \cite{xie2024} propose the first size- and state-aware dispatching policy called CARD (Controlled Asymmetry Reduces Delay), which keeps all but one of the queues short, then routes as few jobs as possible to the one long queue that provably minimizes mean delay in heavy traffic.
Finally, it is well known that \ac{SRPT} minimizes the job response time in a single server system, within the class of size-based, pre-emptive policies.
Moreover, multi-stage systems offer a promising solution for optimizing resource allocation by categorizing tasks based on their service times, improving performance in dynamic and diverse environments. 
However, the success of these approaches largely depends on the characteristics of the workloads and system configurations. 
In real-world scenarios, for example, the service times of incoming tasks are often unknown, which makes size-based policies hard to implement.

This paper examines the performance of various dispatching and scheduling algorithms in large-scale cluster systems, leveraging real-world data from Google's Borg clusters \cite{Tirmazi2020, Verma2015}.
We assess how these algorithms impact response times and system slowdowns across different load conditions and analyze the influence of parallelism on performance. 
Additionally, we propose a novel two-stage dispatching system tailored for managing highly diverse workloads, achieving notable efficiency gains compared to traditional single-stage methods.

This research contributes to the ongoing efforts to optimize resource allocation in large-scale distributed systems, offering valuable insights for improving the performance of cloud computing and high-performance clusters.
More in depth, the specific contributions of this paper are as follows:
\begin{itemize}
	\item We assess the role of parallel servers, by keeping the overall computational power budget fixed, showing that simple, suboptimal policies can achieve as good performance as known optimal ones (\emph{break down correlations and improve performance by means of parallel servers rather than complex policies}).
	\item We give evidence that a suitable system architecture (two-stage server system) can outperform single-stage one, in spite of the latter using much more powerful policies than the former system (\emph{buy performance gain with design of system architecture, rather than complex policies}).
\end{itemize}

The remainder of this paper is organized as follows: 
\cref{sec:LitRev} reviews related work on dispatching and scheduling algorithms for large-scale systems. 
\cref{sec:dataset} details the dataset and key metrics used in our simulations. 
In \cref{sec:twostage}, we introduce our proposed approach. 
\cref{sec:SysDesign} describes our experimental setup, while \cref{sec:simresults} analyzes the simulation results. 
Finally, \cref{sec:conclusion} summarizes our findings and discusses potential directions for future research.

%

\section{Background and Related Work}
\label{sec:LitRev}

In recent years, the management of resources in data centers has been extensively studied from multiple perspectives, particularly in terms of determining the optimal amount of computational power and effectively allocating resources to various jobs. 
Researchers have explored various strategies to balance energy efficiency, performance, and scalability, focusing on both the hardware and software layers. 

Optimizing dispatching and scheduling in multi-server environments remains a complex task, with no universally optimal solution identified \cite{Chraibi2023}. 
While significant advances have been made in the analysis and optimization of single-server systems, the multi-server context presents challenges such as heterogeneous resource demand, and load distribution across servers, making it difficult to apply a single solution across different scenarios \cite{HarcholBalter2021}. 

The optimality of a dispatching policy heavily depends on the information available to the dispatcher. 
When job sizes are unknown but server states (e.g., queue lengths or workloads) are accessible, policies like \ac{RR} \cite{Zhen1994,Zhen1998}, \ac{JSQ} \cite{Richard1978}, and \ac{LWL} \cite{Osman2013} are proven to be optimal by directing jobs to the least-loaded servers. 
On the other hand, when only job sizes and their distribution are known, the \ac{SITA} policy is considered optimal \cite{FENG2005}. 
However, these assumptions often do not align with the complexities of real-world workloads.

In \cite{Caglar2022} and \cite{Singh2021}, the authors make a strong case for enhancing energy efficiency by utilizing fewer, more powerful servers. 
While traditional dispatching and scheduling methods often indicate that a higher number of servers can reduce mean response time, our innovative two-stage approach shows superior performance even with fewer servers that have higher computational power. 
By applying standard algorithms across both stages, the system effectively optimizes workload distribution and resource management. This results in better performance compared to conventional methods.
Furthermore, consolidating tasks onto fewer powerful servers reduces both energy consumption and operational overhead, as these servers can efficiently handle more complex workloads, thereby avoiding the inefficiencies associated with managing a larger fleet of lower-powered servers.

A large heterogeneous server cluster is considered in \cite{Choudhury2021} by means of a discrete-time (slotted) model, assuming Bernoulli arrivals and Geometric service times.
Under limited knowledge of server capacity and with delayed queue length information, the authors argue that there is no point in resorting to complex policies such as \ac{JSQ}, whereas weighted \ac{RR} grabs most of the potential performance.
The emphasis of this work is on the optimization of mean system response time under limited knowledge of server capacity, with a bandit-based exploration.

In our work, we explore the effectiveness of large-scale cluster sizes in taming real-world workload variability.
Inspired by  \cite{Caglar2022,Singh2021}, we then consider a two-stage architecture for the server cluster, to reap as good performance as with large-scale systems, while not requiring a very large number of servers.
Also, following \cite{Choudhury2021}, but using real-world workload measurement, we show that \ac{RR} is still competitive with respective to more sophisticated policies.
Instead of using dynamic learning systems, we show that a substantial improvement in performance is achievable by playing with system architecture.

In our investigation, we found that while the heuristic algorithm we tested may not be globally optimal, it outperformed some of the best-known policies in specific conditions. 
This observation aligns with recent findings, such as those discussed in \cite{Efrosinin2023}, which highlight the difficulties of achieving optimal scheduling in complex, heterogeneous systems. 
Additionally, \cite{Grosof2022} explored optimal scheduling under heavy traffic in multi-server environments and emphasized the challenge of balancing workloads across servers while maintaining efficiency. 
Their findings further support the notion that, while heuristic methods may provide substantial improvements, achieving a universally optimal scheduling strategy remains elusive. 

Our research addresses the role of server cluster size and architecture under real large-scale workload data, demonstrating that a two-stage system design can significantly enhance performance, even when using simpler, suboptimal policies. 
This approach emphasizes that optimizing system architecture, rather than relying solely on complex scheduling policies, can achieve superior results, providing valuable insights into the effective allocation of resources in large-scale distributed environments.

%

\section{Workload Dataset Description}
\label{sec:dataset}

The section describes the Google comprehensive traffic dataset released in 2020, which captures user and developer activity measured over one month in 2019.
This dataset provides detailed data on users' requests of Google's computational resources, such as \ac{CPU}, across eight data centers located in America, Europe, and Asia, referred to by Google as ``Borg cells.'' 
Traffic measurements from each Borg cell consist of five tables with information about users' resource requests, the machines processing these requests, and the time evolution of each request through its processing states in Borg's scheduler.

Users submit jobs, which may consist of one or more tasks, known as instances. 
Each instance requires a specific amount of \ac{CPU} time to execute successfully. 
The dataset provides these quantities in units based on the so called \ac{GNCU}.
The \ac{GNCU} is the computational power of a default machine. 
\ac{CPU} times provided for each task assume the server computational power equals 1 \ac{GNCU}.

The life cycle and event types associated with jobs and tasks are crucial to understanding their execution.
After a job is submitted, its tasks may either be processed immediately or placed in a queue, depending on the system's load. 
Upon completion, tasks are categorized based on their outcome. 
Google classifies successfully completed tasks as ``FINISH''.
Those that do not complete successfully are labeled as ``KILL,'' ``LOST,'' or ``FAIL,'' depending on the specific reason for the failure. 
Further details can be found in the documentation \cite{wilkes2019clusterdata}.

After conducting a detailed analysis, we extracted the key information necessary for managing resources in large-scale clusters \cite{yildiz2024}. 
Our focus was limited to tasks that were successfully completed (``FINISH'' tasks), as incomplete tasks do not provide reliable insights into the resources required for completion.

Google’s original dataset includes essential details for each task, such as job and task identifiers, the amount of \ac{CPU} time utilized (averaged over 300-second intervals), and randomly sampled 1-second snapshots of \ac{CPU} usage within each time window. 
Using this data, we reconstructed a workload description with the following features for each task:\footnote{The dataset used in this study, encompassing data from all eight Google data centers, is publicly available at \textcolor{blue}{\url{https://github.com/MertYILDIZ19/Google_cluster_usage_traces_v3_all_cells}}.}

\begin{itemize}
    \item $A_{ij}$: The arrival time of task $i$ belonging to job $j$.
    \item $C_{ij}$: The \ac{CPU} time required to execute task $i$ of job $j$ on a standard processor with a computational capacity of 1 \ac{GNCU}, representing the service time for the task.
\end{itemize}

Google's dataset provides performance metrics for these ``standard'' processors, enabling consistent workload analysis. 
In this study, we focus on a simplified task representation, considering only the arrival times and the required \ac{CPU} time for each task. This abstraction allows for a more streamlined evaluation of system performance while maintaining sufficient detail for meaningful insights.


\section{Framework for the Two-Stage Dispatching and Scheduling System}
\label{sec:twostage}

The two-stage dispatching system is designed to address the challenges of managing highly heterogeneous workloads in large-scale multi-server environments. 
The system separates tasks based on their service time, ensuring efficient resource utilization by isolating short tasks ("mice") from long tasks ("elephants"). 
This architectural approach aims to improve performance without relying on complex dispatching policies.

\begin{figure}[ht]
\vspace{-0.325cm}
\centering
\includegraphics[width=0.9\columnwidth]{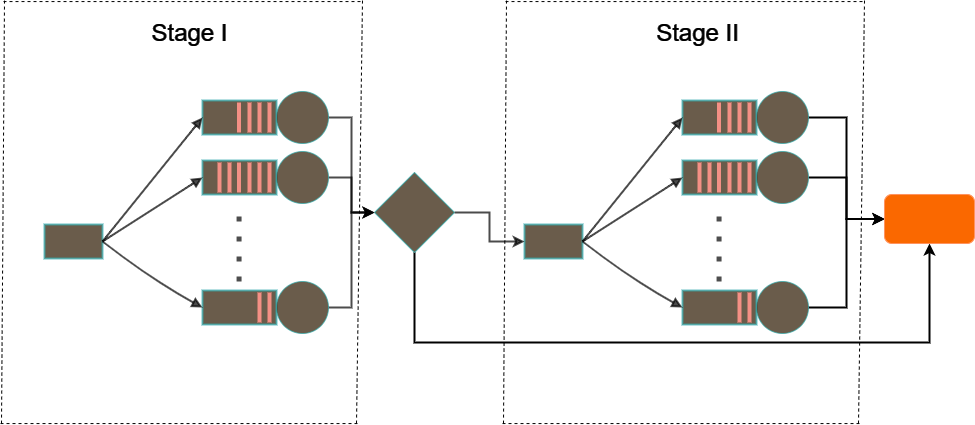}
\caption{Layout of the Two-stage Multi-Server System.}
\label{fig:Flowch}
\end{figure}

The layout of the two-stage system is illustrated in \cref{fig:Flowch}. 
It consists of two clusters of servers: the first-stage cluster with $N_1$ servers and the second-stage cluster with $N_2 = N - N_1$ servers, where $N$ is the total number of servers. 
Tasks arriving at the system are dispatched to the first-stage servers. 
If a task is completed within a predefined threshold $\theta$, it exits the system.
Otherwise, it is transferred to the second-stage cluster for further processing. The operation of the system is formalized in Algorithm~\ref{alg:twostage}.

\begin{algorithm}[ht]
\caption{Two-stage Dispatching and Scheduling System}
\label{alg:twostage}
\begin{algorithmic}[1] 
\State \textbf{Input:} Task set $T = \{t_1, t_2, \dots, t_n\}$, Threshold $\theta$, Stage\_1 Servers $N_1$, Stage\_2 Servers $N_2$
\State \textbf{Initialize:} RR dispatchers and FCFS scheduling for servers for Stage\_1 and Stage\_2

\Procedure{Stage\_1 Dispatching and Processing}{}
    \For{each arriving task $t_i \in T$}
        \State Dispatch $t_i$ to Stage\_1 server $N_1$ using RR
        \While{$t_i$ is not completed}
            \State Process $t_i$ on $N_1$ using FCFS
            \If{$t_i$ is completed before $\theta$}
                \State Exit system
            \Else
                \State Transfer $t_i$ to Stage\_2 dispatcher
                \State \Call{Stage\_2 Processing}{$t_i$}
            \EndIf
        \EndWhile
    \EndFor
\EndProcedure

\Procedure{Stage\_2 Processing}{$t_i$}
    \State Dispatch $t_i$ to Stage\_2 server $N_2$ using RR
    \While{$t_i$ is not completed}
        \State Process $t_i$ on $N_2$ using FCFS
    \EndWhile
    \State Exit system
\EndProcedure
\end{algorithmic}
\end{algorithm}

This system design employs simple dispatching and scheduling policies to minimize implementation complexity. 
\ac{RR} is used for dispatching, and \ac{FCFS} is employed for scheduling at both stages. 
Despite its simplicity, the two-stage architecture effectively manages heterogeneous workloads through task separation and resource specialization.

The proposed framework emphasizes architectural design over sophisticated policies. 
By demonstrating the effectiveness of this design, the system showcases a practical and scalable solution for real-world distributed environments.


\section{Simulation Setup}
\label{sec:SysDesign}

This section describes the simulation setup for evaluating multi-server systems using Google's workload dataset. 
Simulations are run by feeding the multi-server system with one day worth of workload.

Considered performance metrics are:
\begin{enumerate}
  \item \ac{MRT}, $R$: the average across all jobs of the time elapsing since when the first task of the job arrives, until when the last task of the job is completed.
  \item \ac{MJS}, $S$: the average across all jobs of the ratio between the job response time and the maximum among service times required by all tasks belonging to the job.
\end{enumerate}

Two different scenarios are defined for simulations: the single-stage service system in \cref{subsec:expsetup1phase}, and the two-stage one in \cref{subsec:expsetup2phase}.

\subsection{Single-stage multi-server system}
\label{subsec:expsetup1phase}

The workload built out of Google's data is fed to a multi-server system, comprising $N$ parallel, fully accessible and equivalent servers.
Each server is configured with $\mu$ \acp{GNCU} and adopts \ac{FCFS} policy.
Hence, the service time of task $i$ of job $j$ is given by $X_{ij} = C_{ij}/\mu$.
The overall average utilization coefficient of the multi-server system (briefly referred to as load) is
\begin{equation}
\label{eq:definitionofrho}
\rho = \frac{ \lambda \overline{X} }{ N }
\end{equation}
where $\lambda$ is the mean task arrival rate computed as
\begin{equation}
\label{eq:definitionoflambda}
\lambda = \frac{ \max_{ i,j }{ A_{ij} } - \min_{ i,j }{ A_{ij} } }{ \sum_{ j = 1 }^{ J }{ K_j } }
\end{equation}
and $\overline{X}$ is the average service time, given by
\begin{equation}
\label{eq:definitionofaverageX}
\overline{X} = \frac{ \sum_{ j = 1 }^{ J }{ \sum_{ i = 1 }^{ K_j }{ X_{ij} } } }{ \sum_{ j = 1 }^{ J }{ K_j } } = \frac{ 1 }{ \mu } \, \frac{ \sum_{ j = 1 }^{ J }{ \sum_{ i = 1 }^{ K_j }{ C_{ij} } } }{ \sum_{ j = 1 }^{ J }{ K_j } } = \frac{ \overline{C} }{ \mu }
\end{equation}

Note that $\lambda$ and $\overline{C}$ (the average service time when servers are endowed with one \ac{GNCU}) are fixed numbers, once the Google dataset workload trace is given.
The equations above imply that
\begin{equation}
\label{ }
\rho = \frac{ \lambda \overline{C} }{ \mu N }
\end{equation}
Then, we define two kinds of analysis:
\begin{enumerate}
	\item we vary the load $\rho$ on the system by scaling $\mu$, for a fixed number of servers $N_0$.
	\item we vary the number of servers $N$ by scaling $\mu$, for a fixed load $\rho_0$.
\end{enumerate}

Dispatching across servers is managed using one of several possible policies: \ac{RR}, \ac{JIQ}, and \ac{LWL}.
These policies have been chosen so as to compare performance achieved by increasingly more complex policies: from the simple stateless \ac{RR} to the stateful, but simple pull policy \ac{JIQ}, up to the anticipative policy \ac{LWL}.

\subsection{Two-stage multi-server system}
\label{subsec:expsetup2phase}

The two-stage system simulation leverages real-world workload traces from Google’s Borg clusters to evaluate performance under diverse conditions. 
The system consists of $N$ servers, each with computational power $\mu$ \ac{GNCU}, divided equally between two stages: $N_1 = N_2 = N/2$. 
The total computational budget is maintained constant at $N \mu = \lambda \overline{C} / \rho_0$, where the system load $\rho$ is fixed at $\rho_0$, and $\mu$ is adjusted to satisfy $\lambda \overline{C} / (\mu N) = \rho_0$.

Tasks are dispatched to the first stage using the round-robin (RR) policy. 
If a task is completed within a predefined threshold $\theta$, it exits the system.
Otherwise, it is transferred to the second stage for further processing, where both stages employ first-come, first-served (FCFS) scheduling. 
Different values of the threshold $\theta$ are tested, as outlined in \Cref{tab:parameters}.

The performance metrics used for evaluating the two-stage system are consistent with those applied in the single-stage system, ensuring comparability. 
This design facilitates an analysis of the effects of task separation and resource specialization on system performance.

By varying the number of servers $N$ and computational capacity $\mu$, this simulation investigates the scalability and efficiency of the two-stage system. 
The setup provides insights into how parallelization and architectural design influence performance in heterogeneous workload scenarios.


\section{Simulation Results}
\label{sec:simresults}

This section presents the simulation results, according to the system set-up defined in \cref{sec:SysDesign}.
We evaluate how dispatching policies, system architecture, and parallelism affect performance metrics like mean job response time and mean job slowdown, highlighting the benefits of the two-stage system over single-stage configurations.

\cref{tab:parameters} summarizes the key parameters used in our simulations.
Notably, in our experiments, the number of servers in each stage of the two-stage system was kept the same for both stages. 
By maintaining a constant overall capacity while varying the number of servers ($N$) and the processing rate per server ($\mu$), under the constraint that the product $N \mu$ be fixed, according to the fixed utilization coefficient $\rho_0$, we systematically examined how changes in parallelism and architectural design influence system performance.

\begin{table}[t]
\small
\centering
\caption{Simulation parameters}
\label{tab:parameters}
\resizebox{\columnwidth}{!}{%
\begin{tabular}{|l|l|}
\hline
\textbf{Parameter}            & \textbf{Value/Description}                   \\ \hline
Number of Servers, $N$        & [2, 3000]        \\ \hline
Ratio of number of servers    & \\
in two-stage multi-server     & 1\\
system, $N_{1}$/$N_{2}$       & \\ \hline

Server Processing Rate, $\mu$ & Computed based on fixed overall capacity     \\ \hline
System load, $\rho$           & [0.5; 0.9]               \\ \hline
Target fixed load, when varying $N$	& 0.6	 	\\ \hline
Threshold, $\theta$           &  \{0.5, 1, 2, 5, 10, 20, 50, 75, 100, 200\} sec    \\ \hline
Dispatching Policies          & RR, JIQ, LWL                           \\ \hline
Scheduling Policy             & FCFS                                         \\ \hline
Performance Metrics           & Mean Response Time, Slowdown          \\ \hline
\end{tabular}%
}
\vspace{-0.375cm}
\end{table}

These parameters were chosen to reflect realistic conditions often found in large-scale computing environments. 
For instance, varying the number of servers allows us to assess the scalability of different dispatching policies under increasing parallelism while adjusting the processing rate per server provides insights into the trade-offs between server capacity and load distribution. 

Performance metrics defined in \cref{sec:SysDesign} are used to quantify the effectiveness of each configuration, enabling a direct comparison between single-stage and two-stage systems.
Our evaluation focused on three dispatching algorithms, namely \ac{RR}, \ac{JIQ}, and \ac{LWL}, combined with \ac{FCFS} scheduling within all servers.
The simulation results highlight the critical impact of parallelism and dispatching strategies on system performance.

\begin{figure}[t]
\centering
\includegraphics[width=0.68\columnwidth]{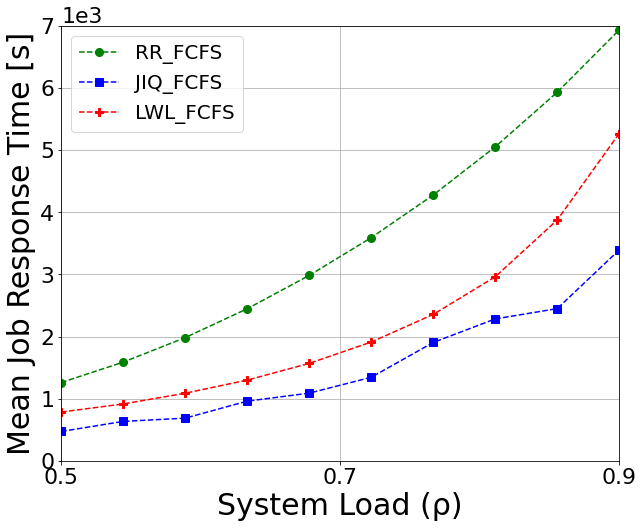}
\caption{\ac{MRT} as a function of system load using different job dispatching and scheduling algorithms.}
\label{fig:Util_MRT}
\vspace{-0.325cm}
\end{figure}

\cref{fig:Util_MRT} shows how \ac{MRT} scales with system utilization coefficient $\rho$ for different dispatching policies. 
\ac{RR} performs poorly, especially under high loads, not surprisingly, given that it is the simplest dispatching policy.
It does not account for the state of the server, nor does it assume knowledge of task sizes.
As a consequence, \ac{MRT} yields strongly worse \ac{MRT} values than \ac{JIQ} and \ac{LWL}.

In contrast, more sophisticated strategies such as \ac{JIQ} and \ac{LWL} effectively leverage parallelism to balance the load across servers. 
\ac{JIQ} assigns tasks to idle servers, ensuring workload distribution without requiring prior knowledge of task sizes. 
However, under heavy load with a fixed number of servers, \ac{JIQ} is known to be suboptimal.

\begin{figure}[t]
\centering
\includegraphics[width=0.68\columnwidth]{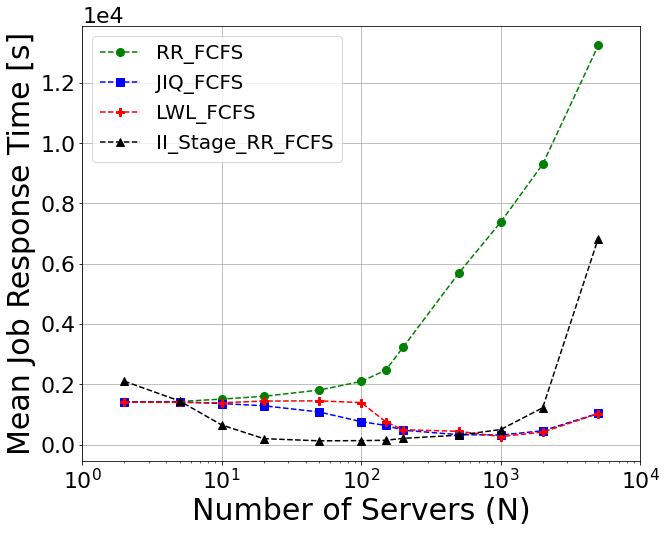}
\caption{\ac{MRT} as a function of the number of servers, demonstrating the performance impact of increasing parallelism.}
\label{fig:Nse_MRT}
\vspace{-0.325cm}
\end{figure}

\Cref{fig:Nse_MRT} illustrates the \ac{MRT} as a function of the number of servers for a given mean load $\rho_0 = 0.6$. 
As expected, \ac{RR} performs the worst, showing no benefit from increased server parallelism. 
Instead, the \ac{MRT} for \ac{RR} increases with the number of servers $N$.

A more nuanced behavior is observed with \ac{JIQ} and \ac{LWL}. 
Initially, the \ac{MRT} remains nearly constant. 
For moderate values of $N$ (in the range of hundreds), the \ac{MRT} decreases, benefiting from improved parallelism. 
However, as $N$ grows beyond several thousand, the \ac{MRT} starts to increase again.

This rise in \ac{MRT} for large $N$ is due to the inverse relationship between the number of servers and their individual processing rate $\mu$, under the constraint $N \mu$. 
As $N$ increases, $\mu$ decreases, resulting in slower processing speeds per server. 
While additional servers increase the likelihood of finding an idle server, the reduced speed of each server causes tasks to be processed more slowly overall. 
Consequently, the \ac{MRT}, which is inversely proportional to $\mu$, eventually grows unbounded as $N$ approaches infinity.

The local minimum of \ac{MRT} represents the optimal trade-off between server speed and parallelism. 
At this point, the benefits of reducing workload correlations through parallelism balance out the negative impact of slower server speeds. 
Beyond this point, the diminishing returns of parallelism and the effect of decreasing $\mu$ dominate, leading to an overall increase in \ac{MRT}.

Another interesting point is that \ac{JIQ} tends to achieve the same performance as \ac{LWL}, even though the latter is exploiting the knowledge of task sizes and of server backlogs.
Such detailed information, often hard to get in real systems, appear to be superfluous for large-scale cluster sizes.

\begin{figure}[t]
\centering
\includegraphics[width=0.68\columnwidth]{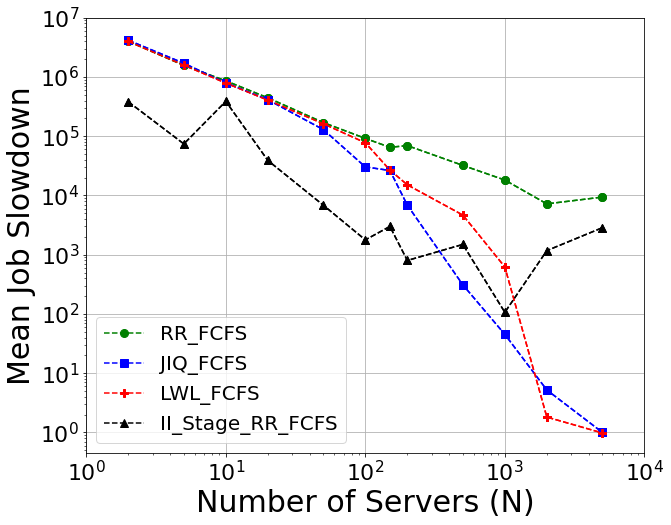}
\caption{Job Slowdown as a function of the number of servers, illustrating the scalability benefits of parallelism.}
\label{fig:Nse_SD_Log}
\vspace{-0.325cm}
\end{figure}

\cref{fig:Nse_SD_Log} illustrates the behavior of average job slowdown as the number of servers grows.
With increasing server counts, all dispatching algorithms show improvements in performance.
These results confirm the inherent strengths of parallelism. 

\begin{figure}[t]
\centering
\includegraphics[width=0.68\columnwidth]{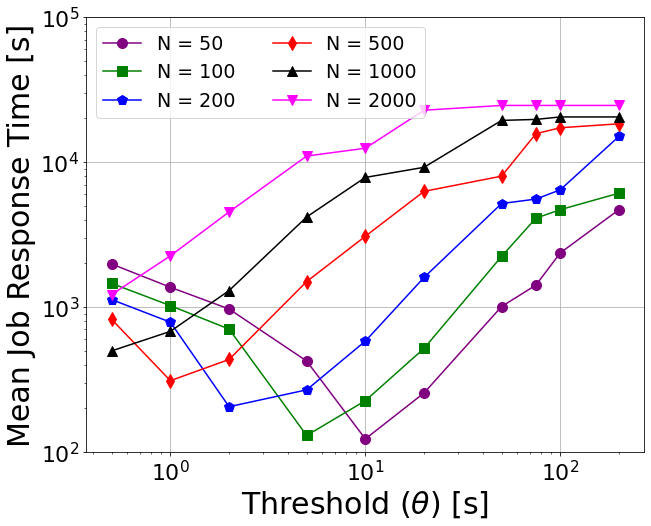}
\caption{\ac{MRT} of jobs for different server counts and threshold values in the two-stage system, showcasing its superior performance.}
\label{fig:II_phase_MRT_TH}
\vspace{-0.325cm}
\end{figure}

Our proposed two-stage system stands out as the top performer, outperforming even the most sophisticated dispatching policies while employing the simplest \ac{RR} strategy. 
As illustrated in \cref{fig:Nse_MRT}, the two-stage system consistently achieves lower \ac{MRT} across different thresholds and server configurations. 
Remarkably, it delivers better \ac{MRT} with just 20 servers than a single-stage system with 1000 servers, even when using the best dispatching policy among those considered here.
By isolating short tasks ("mice") from longer ones ("elephants"), the two-stage system minimizes delays for smaller tasks, leading to more efficient resource utilization and significantly improved response times. 
However, when the number of servers is relatively large, the two-stage system loses again single stage one systems. 
This is because the processing speed of each server is very low, causing tasks to get stuck in the first stage, while subtracting server to the second-stage to deal with bigger tasks.
The two-stage system's performance is particularly noteworthy given its use of the \ac{RR} dispatching policy, which is typically the least effective in single-stage scenarios. 
Despite this, our approach outperforms \ac{LWL} with \ac{FCFS}, demonstrating the profound impact of architectural design on system efficiency.

\cref{fig:II_phase_MRT_TH} plots the \ac{MRT} as a function of the service time threshold of the first stage, $\theta$, for several values of $N$ and a fixed value of the average utilization factor $\rho_0$.
The optimal threshold value that minimizes the \ac{MRT} is identified from these curves.


\section{Conclusion}
\label{sec:conclusion}

In this paper, we evaluated the performance of various dispatching and scheduling algorithms in large-scale cluster systems using real-world workload measurements from Google's Borg clusters. 
Our proposed two-stage dispatching and scheduling system showed significant improvements in managing heterogeneous workloads, outperforming traditional single-stage models, including those employing powerful dispatching strategies like \ac{LWL}. 
Overall, these results emphasize the importance of parallelism and architectural design of the server cluster system.
Playing with those levers, it is possible to achieve as good performance (or even better than) those achievable with complex dispatching policies.

For future work, we plan to extend our approach by exploring systematic optimization of the parameters of the two-stage system (e.g., the threshold $\theta$, the size of the first stage $N_1$, and possibly the capacity of servers in the first and second stage) is another interesting line of development.
Finally, trying to gain more insight into system architecture optimization employing, possibly approximate, models is also worth of being pursued.
It is especially interesting to understand up to what point simplifying assumptions on workload statistics, required to define models amenable to analysis, might lose key effects brought about by real-world workload traffic traces provided by Google's measurements.

\bibliographystyle{ieeetr}
\bibliography{references}

\end{document}